\documentclass[aps,twocolumn,showpacs,pre]{revtex4}

\usepackage{graphicx}
\usepackage{dcolumn}
\usepackage{bm}
\usepackage{epstopdf}
\usepackage{amsmath}
\usepackage{multirow}

\begin{document}

\title{Smectic Liquid Crystals in an Anisotropic Random Environment}

\author{Dennis Liang}
\thanks{Present address: Intense Pulsed Neutron Source Division, Argonne 
National Laboratory, Argonne, IL 60439}
\author{Robert L. Leheny}

\affiliation{Department of Physics and Astronomy, Johns Hopkins 
University, Baltimore, MD 21218, USA}

\date{\today}

\begin{abstract}

We report a high-resolution x-ray scattering study of the smectic liquid 
crystal octylcyanobiphenyl (8CB) confined to aligned colloidal aerosil 
gels.  The aligned gels introduce orientational fields that promote 
long-range nematic order while imposing positional random fields that 
couple to the smectic density wave and disrupt the formation of an 
ordered smectic phase.  At low densities of aerosil, the low-temperature 
scattering intensity is consistent with the presence of a topologically 
ordered XY Bragg glass phase that is predicted to form in response to 
such anisotropic quenched disorder.  The observed features of the phase 
include an algebraic decay of the smectic correlations, which is 
truncated at large length scales due to the imperfect nematic order, and 
a power-law exponent that agrees closely with the universal value 
predicted for the XY Bragg glass.  At higher aerosil densities, 
deviations from the XY Bragg glass form are apparent.  At high 
temperature, the scattering intensity displays pre-transitional dynamic 
fluctuations associated with the destroyed nematic to smectic-A 
transition.  The fluctuations obey quasi-critical behavior over an 
extended range of reduced temperature.  The effective critical exponents 
for the correlation lengths and smectic susceptibility differ 
systematically from those of pure 8CB, indicating that coupling of the 
nematic order to the gel suppresses its role in the smectic critical 
behavior.

\end{abstract}


\maketitle

\section{Introduction}

The effect of impurities and quenched disorder in condensed matter 
represents an important problem both because of the ways in which 
disorder can change fundamental properties of a system and because any 
real material will inevitably possess imperfections.  Studies have shown 
dramatic consequences of disorder, such as the destruction of ordered 
phases and the introduction of new exotic ones, deviation from the 
universal behavior around transition points, and sometimes even the 
enhancement of useful material properties.  Liquid crystals, due to 
their soft elasticity, sensitivity to surface interactions, and 
experimental accessibility, provide an excellent opportunity for 
studying the effects of quenched disorder, which can be introduced 
experimentally through confinement of the liquid crystal to random 
porous media.  A particular focus in this area has been the study of 
smectic liquid crystals in silica gels.  In a detailed theoretical 
treatment of smectics in such random environments, Radzihovsky, Toner, 
and coworkers identified two forms of random-field disorder that the 
confinement introduces to the smectic -- random orientational fields 
that couple to the nematic director and random positional fields that 
couple to the smectic density wave~\cite{radz_toner1,
 radz_toner2,radz_toner3,radz_toner4, saundersthesis}. 
One conclusion of their study was that the smectic phase is 
unstable to arbitrarily weak random-field disorder, consistent with 
expectations for a phase that breaks a continuous symmetry in three 
dimensions.  Experimentally, considerable work has been dedicated toward 
investigating smectics in two types of silica gels, chemically rigid 
aerogels~\cite{clark-science,lei,finotello-gel,bellini-aerogel-light} 
and colloidal gels formed from aerosil particles~\cite{park,
finotello-sil,paperI,paperII,Liang,finotello-pre,zhou,Mertelj, theon,germano, 
aerosil1,aerosil2,aerosil3,aerosil4,aerosil5,aerosil6,clegg8S5,clegg8OCB,
Larochelle,8CBrheo}.  The chemical compositions and structures of the 
two gels are very similar, but the hydogen-bonded aerosil gels have the 
potential to access a weaker regime of disorder.  Utilizing a wide range 
of probes, experiments on smectics confined to aerogels and aerosil gels 
have detailed the consequences of the confinement for the nematic and 
smectic phases, supporting the general picture that the gels impose 
random-field disorder that destroys the nematic to smectic transition.

Among the predictions for smectics confined to gels is the potential 
formation of topologically ordered, ``Bragg glass'' phases that are 
distinct from the high-temperature nematic phase by the absence of 
unbound dislocation loops.  Considerable evidence from theory and 
simulation supports the existence of a topologically ordered state with 
algebraic decay of correlations in the three-dimensional (3D) random 
field XY model at low temperature~\cite{giamarchi,huse,fisher}.  
(However, a recent argument based on the functional renormalization 
group approach has questioned the existence theoretically of such a 
phase in 3D XY systems~\cite{tarjus}.)    Since the nematic to smectic-A 
(N-SmA) transition breaks 3D XY symmetry, one might hence expect that 
smectics confined to gels would be good candidates for realizing such an 
``XY Bragg glass'' phase.  Considering this possibility theoretically, 
Radzihovsky and Toner have concluded that the orientational random 
fields coupling to the nematic order and the soft elasticity of the 
smectic make smectics confined to gels distinct from standard random 
field XY systems.  Instead, they have introduced the possibility that  a  
``smectic Bragg glass'' phase,  which has short range correlations and 
is qualitatively different from the XY Bragg glass, might form in 
smectics confined to gels~\cite{radz_toner1}.  However, the theoretical 
argument for the stability of the smectic Bragg glass is inconclusive, 
and experimental evidence for the phase in smectics confined by silica 
gels has been contradictory~\cite{clark-science,
paperI,paperII,clegg8OCB}.  Jacobsen, Saunders, {\it et al.} 
have further argued that smectics confined to uniaxially strained gels 
should fall into the universality class of random field XY systems and 
hence should form an ``XY Bragg glass''  for sufficiently weak 
disorder~\cite{radz_toner4, saundersthesis,toner_prl}.  The key 
difference between isotropic gels and strained gels is the effect of the 
orientational fields.  Specifically, if the strain alters the 
distribution of orientational fields, causing them to suppress nematic 
fluctuations at large length scales, then the smectic in strained gels 
should theoretically exhibit an XY Bragg glass, provided the positional 
random fields are sufficiently weak.  Recent additional theoretical 
predictions regarding the smectic-A to smectic-C transition under 
confinement in strained gels has highlighted the potential for smectics 
in such an anisotropic environment to display exotic phase 
behavior~\cite{toner_prl}.  

In an effort to understand the separate effects of orientational random 
fields and positional random fields on the N-SmA transition and to test 
for the presence of an XY Bragg glass phase, we have conducted an x-ray 
scattering study of smectic octylcyanobiphenyl (8CB) confined to {\it 
aligned} aerosil gels.  The structure of these gels dramatically alters 
the nature of the orientational fields, converting the random 
distribution found in isotropic gels into a sharply anisotropic 
distribution that macroscopically orders the nematic director.  This 
anisotropy, or ``soft axis'', makes the gels a faithful experimental 
realization of uniaxially strained systems considered by Jacobsen, 
Saunders, {\it et al.}~~In this paper, we report an analysis of the x-ray 
scattering results that indicates the presence of an XY Bragg glass 
at low temperature for low densities of aligned aerosil.  The signature 
feature of the XY Bragg glass is a power-law divergent smectic 
scattering peak with a universal exponent.  We note that in a 
preliminary report on our x-ray studies~\cite{Liang}, we concluded that 
8CB confined to aligned aerosil gels did not exhibit such 
characteristics of the XY Bragg glass phase.  As we explain below, by 
adopting a physically better-motivated description of the behavior of 
the thermal critical fluctuations  in smectics confined to aerosil 
gels~\cite{Larochelle}, we find that the previous analysis is not 
satisfactory in accounting for the smectic correlations at low 
temperature.  Instead, when a finite nematic domain size is incorporated 
into the analysis to reflect the imperfect nematic order in the aligned 
gels, comparisons between the x-ray scattering results for low densities 
of aerosil and the predictions for an XY Bragg glass show excellent 
agreement.  A discussion of these findings is given in Sec.~III below.  

Another important benefit of the macroscopic alignment of the nematic 
director by the anisotropic gels is the pronounced enhancement in 
signal-to-background in x-ray measurements relative to measurements on 
smectics in isotropic gels.  Taking advantage of this enhancement, we 
further provide in Sec.~IV a detailed study of the smectic correlations 
at high temperature, which we analyze in terms of dynamic fluctuations 
that grow on approaching the critical region of the destroyed N-SmA 
transition.  This analysis shows that the correlations, as characterized 
by effective critical exponents for the correlation lengths and 
susceptibility, differ from those of pure 8CB even when the correlation 
lengths are considerably smaller than the gel pore size, providing 
evidence that the presence of quenched disorder alters the coupling 
between nematic and smectic order and its influence over smectic 
critical behavior.  When combined with similar results from smectics 
confined to isotropic gels for effective critical exponents for the 
specific heat and smectic order parameter~\cite{paperI,paperII, 
Larochelle}, these findings provide a comprehensive picture of an 
evolution in the N-SmA critical fluctuations toward conventional 3D XY 
behavior with increasing quenched disorder.

\section{Sample Preparation and Characterization}

Pure 8CB undergoes an isotropic to nematic transition at $T_{NI}$ = 
313.98 K and a N-SmA transition at $T_{NA}$ = 306.97 K~\cite{germano}.  
Confinement of the liquid crystal to aligned aerosil gels was 
accomplished through a procedure described previously~\cite{Liang}.  
Type 300 hydrophilic aerosil (DeGussa Corp.)~was dried under vacuum at 
393 K for 24 hours.  The dried powder along with appropriate quantities 
of 8CB (Frinton Laboratories, Inc.)~was dissolved in high-purity 
acetone, and the mixtures were sonicated for at least two hours.  The 
solutions were then heated to 318 K to evaporate the solvent slowly.  
After no visible trace of acetone remained, the samples were heated to 
338 K under vacuum.  The resulting composites consisted of 8CB confined 
to isotropic fractal aerosil gels like those of previous studies 
~\cite{park,finotello-sil,paperI,paperII,Liang,zhou,Mertelj}.  To create 
anisotropic gels, we placed each sample in a 2 Tesla magnetic field and 
cycled the temperature between 308 K (nematic phase) and 318 K 
(isotropic phase) at least 100 times.  Due to the magnetic anisotropy of 
nematic 8CB, the director tends to align parallel to the magnetic field, 
in competition with the random orientational fields created by the gel.  
Due to the elasticity of the nematic and the compliance of the gel, this 
competition results in restructuring of the gel to accommodate the 
magnetic anisotropy.   As a result, the gels acquire a structure in 
which the orientational fields are no longer random but on average 
orient parallel to the aligning direction.  Since the gel continues to 
have a random structure positionally, the positional fields that couple 
to the smectic density wave remain random.  Samples were prepared in 
this manner with aerosil densities ranging from $\rho_s=0.027$ g 
sil/cm$^{3}$ 8CB, 
which is just above the gelation threshold,  to 
$\rho_s=0.10$ g sil/cm$^{3}$ 8CB, which is the upper limit above which 
efforts to align the gels were ineffectual.   

The smectic correlations that form in this anisotropic random 
environment were studied through high-resolution x-ray scattering.  The 
experiments were conducted at the X22A beam line of the National 
Synchrotron Light Source.  The beam line is equipped  with a Ge(111) 
monochromator to select a beam  of 10 keV x-rays.  The measurements were 
performed in transmission geometry.  The beam size was approximately 1 
mm $\times$ 1 mm, and the sample thickness was approximately 1.5 mm to 
match the attenuation length of the 10 keV x-rays.  A triple-bounce 
Si(111) channel-cut analyzer crystal was positioned between the sample 
and a scintillation point detector to achieve high wave-vector 
resolution.

\begin{figure}
\centering\includegraphics[scale=0.45]{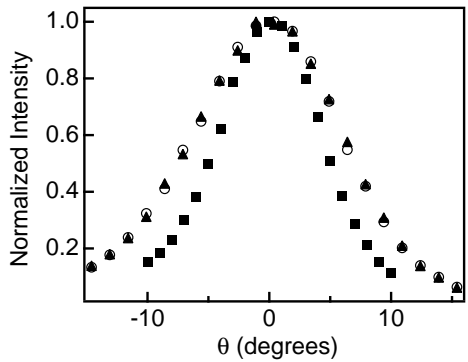}
\caption{Normalized scattering intensity of the smectic peak at fixed 
wave vector magnitude ($q \approx 0.2$~\AA$^{-1}$) as a function of 
sample orientation for 8CB confined by an aligned aerosil gel with 
$\rho_s$ = 0.027 g/cm$^3$ (solid squares) and 0.10 g/cm$^3$ (open 
circles and solid triangles) at 298.15 K.  The two scattering 
intensities for  $\rho_s$ = 0.10 g/cm$^3$ correspond to measurements 
before (open circles) and after (solid triangles) annealing the sample 
in the isotropic phase of 8CB. }
\label{mosaic}
\end{figure}

To determine the degree of anisotropy in the orientational fields 
coupling to the liquid crystal, we characterized the mosaic spread of 
the smectic layer normal by ``rocking curve'' measurements through the 
smectic scattering peak.  Figure~{\ref{mosaic}} shows mosaic scans for 
$\rho_s$ = 0.027 and 0.10 g/cm$^3$ at T=298.1 K, several degrees below 
$T_{NA}$.    
The smectic scattering displays strong azimuthal anisotropy, indicating 
a narrow distribution of smectic layer-normal orientations.  The 
alignment is slightly better (the peak is narrower) for lower density 
gels, consistent with the gels' increased ability to resist 
restructuring with increasing density due to greater in mechanical 
strength~\cite{paperII,8CBrheo}.  We note these measurements were 
performed several weeks after removal of the magnetic field used to 
align the gels; thus, the observed macroscopic alignment is a 
consequence of the gel and not the external field.  As also shown in 
Fig.~{\ref{mosaic}}, the quality of the nematic alignment displays 
negligible change in response to temperature excursions into the 
isotropic phase, demonstrating that the anisotropic gel structure is 
stable against thermal fluctuations and sufficiently robust to re-align 
the nematic director on subsequent cooling.  Indeed, samples are 
observed to remain aligned after repeated heating into the isotropic 
phase and after storage for several months.  Thus, the random 
orientational fields of the isotropic gel have been converted into 
fields with long-range order that align the nematic director 
macroscopically.  

\section{Smectic Correlations:  Evidence for an XY Bragg Glass}

Figure 2(a) displays the low-temperature x-ray scattering intensity for 
$\rho_s$ = 0.027 g/cm$^3$ as a function of wave vector parallel to the 
alignment direction ({\it i.e.}, the direction of the magnetic field 
used to prepare the gel), $q_z$, and perpendicular to it,  $q_x$.  In 
the alignment direction the intensity displays a pronounced peak 
corresponding to the smectic scattering, while in the perpendicular 
direction the intensity shows no evidence of smectic scattering, 
consistent with the strong anisotropy implied by Fig.~1.  We ascribe the 
scattering intensity along $q_x$ entirely to the gel structure and note 
that an essentially identical background intensity contributes to the 
scattering along $q_z$.  In order to isolate the smectic scattering from 
this background, we subtract the intensity along $q_x$ from the 
intensity along $q_z$.  This procedure assumes that the anisotropy in 
the gel structure does not affect the scattering from the gel in the 
wave-vector range of the smectic peak.  As described previously, small 
angle scattering measurements support this assumption~\cite{Liang}.  
Nevertheless, the observed smectic signal greatly exceeds this 
background contribution, by a factor of more than a factor of $10^4$ in 
Fig.~2(a), and this strong signal-to-background, which results from the 
azimuthal focusing by the aligned orientational fields, enables the 
detailed examination of the smectic correlations described below.  
Figure 2(b) displays the results for the scattering intensity along 
$q_z$ from Fig.~2(a) with the background subtracted.  Also in the figure 
is the resolution function determined from the profile of the direct x-ray 
beam.  The inset to Fig.~2(b) compares the scattering to the 
resolution in the region of the peak.  The peak is clearly broader than 
the resolution even for this lowest aerosil density at low temperature.  
\begin{figure}
\centering\includegraphics[scale=0.45]{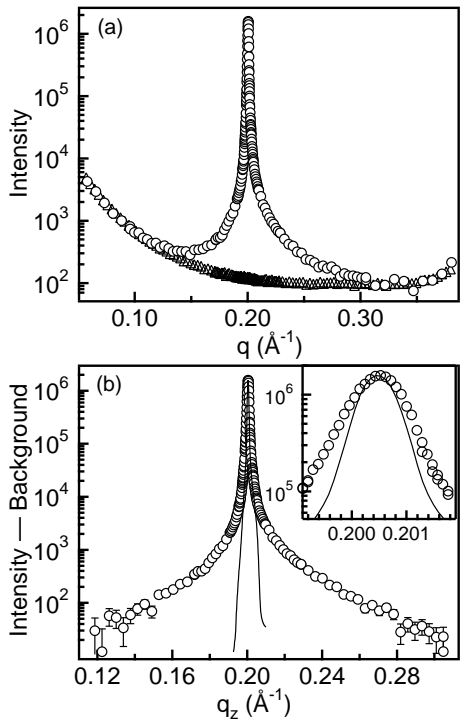}
\caption{(a) X-ray scattering intensity for 8CB confined by an aligned 
aerosil gel with $\rho_{S}=0.027$ g/cm$^{3}$ at 296.15 K as a function 
of wave vector parallel to the gel's alignment direction, $q_z$ 
(circles) and perpendicular to this direction, $q_x$ (triangles).  (b) 
Scattering intensity along $q_z$ with the intensity along $q_x$ 
subtracted to isolate the smectic scattering.  The solid line in (b) is 
the instrumental resolution obtained from the incident beam profile.  
The inset to (b) shows the peak region on an expanded scale.}
\label{peak_bg}
\end{figure}

As a first effort to characterize the smectic order, we consider a 
correlation function based on the random-field model that has been used 
previously to describe the short-range smectic correlations observed 
under confinement in isotropic aerosil 
gels~\cite{park,paperI,clegg8OCB,clegg8S5}.  
Specifically, we model the smectic peak with the two-component line 
shape  
\begin{eqnarray}
    I({\bf q}) = 
    \frac{\sigma_{1}}{1+(q_{\|}-
q_{0})^{2}\xi_{1\|}^{2}+q_{\bot}^{2}\xi_{1\bot}^{2}+c_{1}q_{\bot}^{4}\xi
_{1\bot}^{4}} \nonumber  \\
    +  \frac{a_{2}(\xi_{2\|}\xi_{2\bot}^{2})}{(1+(q_{\|}-
q_{0})^{2}\xi_{2\|}^{2}+q_{\bot}^{2}\xi_{2\bot}^{2}+
    c_2q_{\bot}^{4}\xi_{2\bot}^{4})^{2}} 
\label{ocko_ocko2}
\end{eqnarray} 
where $q_{\|}$ and $q_{\bot}$ are the wave vectors parallel and 
perpendicular to the smectic layer normal, respectively, and $q_{0}$ is 
the ordering wave vector.  The first term in $I(\bf{q}$) is an 
anisotropic Lorentzian with fourth-order correction that describes 
critical dynamic fluctuations on approaching the N-SmA transition in 
pure liquid crystals.  $\sigma_1$ is the smectic susceptibility, and 
$\xi_{1\|}$ and $\xi_{1\bot}$ are the correlation lengths of the 
fluctuations in the directions parallel and perpendicular to the nematic 
director, respectively.  The second term, which has a shape that is 
proportional to the square of the thermal fluctuation term, describes 
static short-range fluctuations due to the random fields, which are 
characterized by the correlation lengths $\xi_{2\|}$ and $\xi_{2\bot}$.  
Previous studies of smectics in isotropic aerosil gels have found that 
the correlations display two temperature regimes.  At low temperature, 
$a_2>0$, and the amplitude of the static fluctuation term rises from 
zero with decreasing temperature.  At higher temperature, $a_2=0$, and 
the scattering profile is similar to that of the nematic phase of pure 
8CB, where pre-transitional smectic critical fluctuations 
dominate~\cite{paperI,clegg8OCB,Larochelle}.   

\begin{figure}
\centering\includegraphics[scale=0.45]{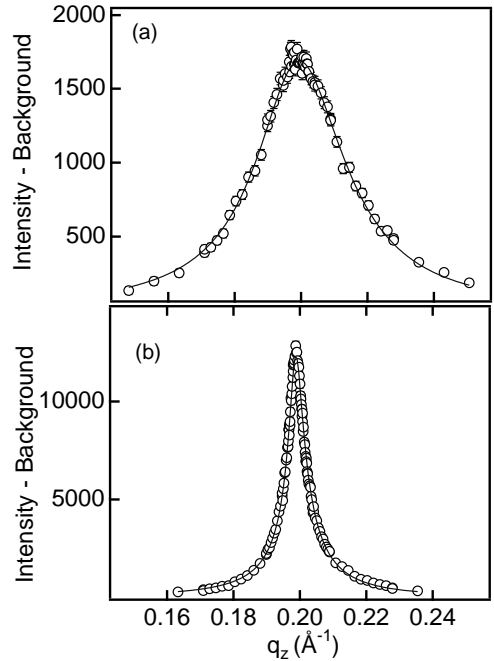}
\caption{Background-subtracted scattering intensity of the smectic peak 
for 8CB confined to an aligned aerosil gel with $\rho_s=0.10$ g/cm$^{3}$ 
at (a) 309 K and (b) 306 K.  The solid lines are the results of fits to 
Eq.~(\ref{ocko_ocko2}) with $a_2 = 0$.} 
\label{highT}
\end{figure}

Figure~{\ref{highT}} shows the results of fits to Eq.~(\ref{ocko_ocko2}) 
with $a_2=0$ at two temperatures in this high-temperature regime for 8CB 
confined by an aligned gel with $\rho_s = 0.10$ g/cm$^{3}$.  
To perform fits to Eq.~(\ref{ocko_ocko2}), we integrate numerically over 
the distribution of layer normal orientations, as determined by rocking 
curves like those shown in Fig.~\ref{mosaic}, and convolve with the 
instrumental resolution.  In this analysis $\xi_{1\|}$ and $\xi_{1\bot}$ 
are treated as independent parameters, while the amplitude of the 
fourth-order correction term, $c$, is treated as a function of 
$\xi_{1\bot}$, with $c(\xi_{1\bot})$ set by the behavior in pure 
8CB~\cite{ocko,J_theon,davidov}.  (Allowing $c$ to vary as an additional 
free parameter leads to significant scatter in the fit parameters.)  As 
Fig.~{\ref{highT}} illustrates, the dynamic term in 
Eq.~(\ref{ocko_ocko2}) provides an excellent description of pre-
transitional smectic fluctuations in 8CB confined to aligned aerosil 
gels at high temperatures.  This agreement allows us to study in detail 
the pseudo-critical behavior of 8CB with anisotropic quenched disorder, 
as described in Sec.~IV below.

Turning to low temperature, the dynamic fluctuation term in Eq.~(1) 
alone cannot describe the measured lineshape.  As mentioned above, 
previous analysis of the smectic correlations under confinement in 
isotropic aerosil gels~\cite{park,paperI,clegg8OCB,clegg8S5} have 
modeled the low-temperature scattering by including the contribution in 
Eq.~(1) from static fluctuations ($a_2 >0$).   In most cases, this 
analysis has assumed that $\xi_{2\|}=\xi_{1\|}$ and 
$\xi_{2\bot}=\xi_{1\bot}$; that is, the correlation lengths in the 
static and dynamic terms were assumed equal.  Typically, this approach 
provided good agreement with the measured line shapes.  Indeed, in a 
preliminary report on 8CB confined to aligned aerosil gels, we found 
that Eq.~(\ref{ocko_ocko2}) with $\xi_{2\|}=\xi_{1\|}$ and 
$\xi_{2\bot}=\xi_{1\bot}$ described the data accurately~\cite{Liang}.
However, this agreement does not necessarily indicate that the analysis 
is capturing the correct physical picture.   As pointed out recently by 
Larochelle {\it et al.}, the results of such analysis lead to trends in 
the smectic susceptibility and correlation lengths that are inconsistent 
with calorimetric studies of smectics confined to aerosil 
gels~\cite{Larochelle}.   Calorimetry measurements on 8CB in isotropic 
aerosil gels at low density ($\rho_{s} < 0.10$ g/cm$^{3}$) show a sharp 
peak in the heat capacity, with power-law temperature dependence 
resembling pre-transitional critical behavior extending over an extended 
range of reduced temperature both above and below an effective 
transition temperature T$^*$~\cite{zhou}.  This behavior would suggest 
that  $\xi_{1\|}$, $\xi_{1\bot}$, and $\sigma_1$ should similarly decay 
as a power law as a function of reduced temperature away from T$^*$.  As 
discussed in Sec.~IV,  for T $>$ T$^*$ the smectic fluctuations in 8CB 
confined to aligned aerosil gels indeed show a quasi-critical increase 
toward T$^*$.  However,  analysis of the x-ray results at low 
temperature using Eq.~(\ref{ocko_ocko2}) with $\xi_{2\|}=\xi_{1\|}$ and 
$\xi_{2\bot}=\xi_{1\bot}$ leads to correlation length and susceptibility 
values that remain large away from T$^*$.  For example, 
$\xi_{\|}\approx$ 3000 \AA~at T$^*-$T = 9 K for 8CB confined to an 
aligned gel with $\rho_{s}=0.10$ g/cm$^{3}$~\cite{Liang}, inconsistent 
with pseudo-critical behavior.  Such large dynamic fluctuations 
persisting to low temperature seem very unlikely.  However, if we 
restrict the dynamic fluctuations to follow a more plausible temperature 
dependence, we find that  Eq.~(1) cannot describe the x-ray lineshapes 
for 8CB in aligned gels at low temperature.  Rather, as we explain 
below, the lineshape predicted for the XY Bragg glass with an 
appropriate cut-off for a finite nematic domain size can very accurately 
describe the measured lineshapes for low densities of aligned aerosil.

\begin{figure}
\centering\includegraphics[scale=0.45]{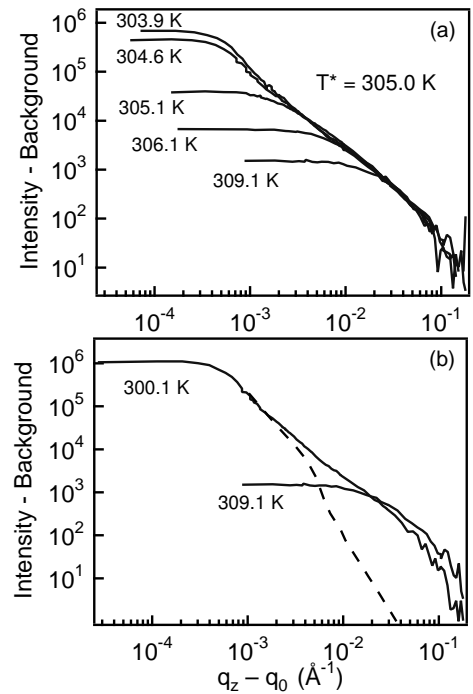}
\caption{Background-subtracted scattering intensity at various 
temperatures as a function of the difference in wave vector from the 
ordering wave vector, $q_0$, for  $\rho_s = 0.042$ g/cm$^3$.  (a)  
intensities at several temperatures ranging from just below T$^*$ = 
305.0 K to several degrees above T$^*$.  (b) intensities at two 
temperatures, 300.1 K and 309.1 K, that are approximately the same 
difference in temperature from T$^*$.  The dashed line in (b) is the 
result of a fitting the intensity at 300.1 K in the peak region 
($(q_z-q_0) < 10^{-3}$  \AA$^{-1}$) to the second term in 
Eq.~(\ref{ocko_ocko2}) ($\sigma_1 = 0$) representing static short-range 
order.}
\label{log_log}
\end{figure}

To examine the situation more closely, we display in 
Figs.~\ref{log_log}(a) and \ref{log_log}(b) log-log plots of the 
scattering intensity versus $(q_z-q_0)$ at various temperatures for 
$\rho_s = 0.042$ g/cm$^3$.  As described in Sec.~IV, the effective 
transition temperature for $\rho_s = 0.042$ g/cm$^3$ is  T$^*$ = 305.0 
K.  As seen in Fig.~\ref{log_log}(a), the intensities at temperatures 
above T$^*$ overlap at large $(q_z-q_0)$.  
This collapse indicates that the intensities away from the peak share 
the same wave-vector dependence, specifically that of the critical 
fluctuation term in Eq.~(\ref{ocko_ocko2}), and further that the ratio 
$\xi_{1\|}^2/\sigma_1$ is a constant independent of temperature 
(implying the critical exponent $(2 - \eta) \approx 2$).  Similarly, for 
the two temperatures just below T$^*$ shown in Fig.~\ref{log_log}(a), 
303.9 K and 304.6 K, the intensities at large $(q_z-q_0)$ collapse onto 
the data for T $>$ T$^*$.  This collapse indicates that critical 
fluctuations continue to dominate the intensity at large $(q_z-q_0)$ at 
temperatures near but below T$^*$.  This finding is not surprising since 
one would expect to have large critical fluctuations in the vicinity of 
a true critical point.  However, for lower temperature, the situation 
becomes more complicated.  Plotted in Fig.~\ref{log_log}(b) is the 
intensity at 300.1 K, approximately 4.9 K below T$^*$, along with the 
intensity at 309.1 K,  approximately 4.1 K above T$^*$.  Not only does 
the low-temperature curve fail to collapse onto the high-temperature 
curve at large $(q_z-q_0)$, but it exhibits considerable excess 
intensity at smaller $(q_z-q_0)$ that cannot be accounted for by either 
term in Eq.~(\ref{ocko_ocko2}).  Specifically, the dashed line in 
Fig.~\ref{log_log}(b) shows the result of fitting the low-temperature 
intensity in the peak region to the static term of 
Eq.~(\ref{ocko_ocko2}) alone ($\sigma_1 = 0$).  While the fit agrees 
adequately with the measured intensity in the immediate vicinity of the 
peak, it deviates significantly at larger $(q_z-q_0)$.  A comparison 
with the high-temperature intensity demonstrates that the dynamic term 
in Eq.~(\ref{ocko_ocko2}) cannot account for these deviations.   Since 
the two intensities plotted in Fig.~\ref{log_log}(b) are for 
temperatures that are approximately the same difference in temperature 
from T$^*$, they correspond to approximately the same reduced 
temperature.  Hence, the pre-transitional critical fluctuations at the 
two temperatures should be similar (assuming the amplitude ratios for 
the correlation lengths and susceptibility are of order one).  For 
scattering from such critical fluctuations, a power law in the intensity 
should be observed only for $( q_z-q_0) > 1/\xi_{1\|}$.  An inspection 
of the line shape for 309.1 K indicates that for this reduced 
temperature this asymptotic range is reached only for $(q_z-q_0) > 4 
\times 10^{-2}$ \AA$^{-1}$.  The extension to lower $(q_z - q_0)$ of the 
power-law behavior at 300.1 K thus indicates that the scattering 
intensity is inconsistent with the small dynamic correlation lengths 
expected at a temperature so far below T$^*$.  In essence, the 
constraint used in the previous analysis~\cite{Liang} that 
$\xi_{2\|}=\xi_{1\|}$ and $\xi_{2\bot}=\xi_{1\bot}$ led to artificially 
large values of $\xi_{1\|}$ and $\xi_{1\bot}$ at low temperature that 
fortuitously approximated the measured line shape but that on further 
inspection cannot be justified as credible behavior for critical 
fluctuations.

The extended power-law tails in the low-temperature scattering peak 
instead suggest the possible presence of the XY Bragg glass phase.  As 
mentioned above, the XY Bragg glass is ideally characterized by a power-
law divergent smectic scattering peak with a universal power-law 
exponent.  As noted by the inset to Fig.~2(a), the measured scattering 
peaks are appreciably broader than the instrumental resolution, which 
precludes such power-law divergence extending to the lowest $(q_z - 
q_0)$.  However, we argue that this discrepancy could be a consequence 
of a large length scale cut-off introduced by the liquid crystal 
ordering in the aligned gels.  Specifically, while the mosaic scans in 
Fig.~1 indicate long-range nematic order, they also show an appreciable 
spread in the smectic layer-normal orientations about the aligning 
direction.  This spread indicates spatial variations in the smectic 
layer normal that are likely set by the length scale characterizing the 
nematic order.   In other words, the liquid crystal in the aligned gel 
has finite-size nematic domains like in the isotropic 
gels~\cite{bellini-sil-PRE}, the difference being that the domains are 
not randomly oriented but rather have a distribution of orientations 
clustered around the aligning direction.  Nevertheless, smectic order 
cannot persist across two misaligned domains, leading directly to a cut-
off in the smectic correlations at low $(q_z - q_0)$.  

Taking such an effect into account, we model the scattering peak at low 
temperature with a form 
\begin{eqnarray}
    I({\bf q}) = P({\bf q}) + H({\bf q})
\end{eqnarray} 
where $P({\bf q})$ is the XY Bragg glass correlation function truncated 
at small $(q_z - q_0)$, and  $H({\bf q})$ accounts for the finite-size 
effect within a Gaussian approximation~\cite{sinha,kaganer}.  The XY 
Bragg glass form is given by~\cite{toner_prl,saundersthesis}
\begin{equation}
    P({\bf q}) = 
    \begin{cases} 
    C((q_{\|}-q_0)^2 + \alpha q_{\bot}^2)^{-\delta/2}  \\  
\text{\hspace{0.5cm} if $((q_{\|}-q_0)^2 + \alpha q_{\bot}^2) > 1/L$,} 
    \\
CL^{\delta} \\  \text{\hspace{0.5cm} if $((q_{\|}-q_0)^2 + \alpha 
q_{\bot}^2) < 1/L$.}
\end{cases}
\end{equation} 
where $\delta$ is a universal, temperature-independent exponent 
predicted to be $\delta = 2.45$~\cite{toner_prl,saundersthesis}, $C$ 
sets the overall amplitude, $\alpha$ is a non-universal constant that is 
expected to be of order unity~\cite{toner_prl}, and $L$ is the size 
scale of the nematic domains.  The finite-size correction has the 
form~\cite{kaganer}
 \begin{equation}
    H({\bf q}) = Ae^{-L^2((q_{\|}-q_0)^2 + \alpha q_{\bot}^2)/4\pi}
\end{equation} 
where the amplitude $A$ is related to $C$ such that at wave vectors  
$((q_{\|}-q_0)^2 + \alpha q_{\bot}^2) < 1/L$ the Gaussian form of 
$H({\bf q})$ dictates the line shape, while for larger wave vectors it 
has negligible effect on the line shape~\cite{sinha}.  To perform fits 
to Eq.~(2), we again integrate numerically over the distribution of 
layer normal orientations, as determined by rocking curves like those 
shown in Fig.~\ref{mosaic}, and convolve with the instrumental 
resolution.  Figure 5 shows the result of a 
\begin{figure}
\centering\includegraphics[scale=0.45]{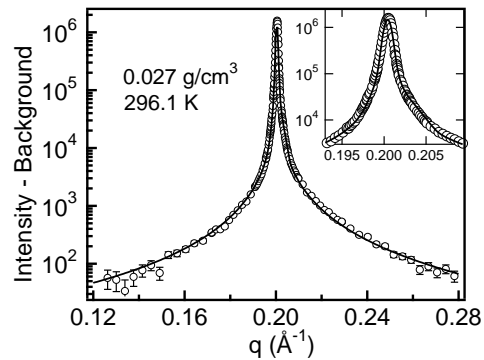}
\caption{Comparison between the background-subtracted scattering 
intensity as a function of wave vector for $\rho_s = 0.027$ g/cm$^3$ at 
296.1 K and the form predicted for an XY Bragg glass with a truncation 
in the correlations at large length scales.  The circles are the 
measured intensity, and the solid line is the result of a fit to 
Eq.~(2).  The inset shows a comparison between the scattering intensity 
and the fit result in the peak region on an expanded scale.  The XY 
Bragg glass form agrees closely with the measured intensity over the 
full range of wave vectors.} 
\label{XYBG0p027}
\end{figure}
fit to Eq.~(2) for $\rho_s$ = 0.027 g/cm$^3$ at T = 296.1 K, the lowest 
measurement temperature.  The figure displays the scattering intensity 
over the full range of wave vectors, while the inset shows the peak 
region on an expanded scale.  The agreement between the measured line 
shape and the XY Bragg glass form is essentially perfect.  Good 
agreement is similarly found for $\rho_s$ = 0.42 g/cm$^3$, as 
illustrated by Fig.~6 which shows the result of a fit using Eq.~(2) to 
the scattering intensity at T = 298.1 K for this aerosil density.  The 
fits for both densities give $L \approx 1$ $\mu$m, in reasonable 
agreement with a value expected based on the nematic correlation lengths 
measured in isotropic aerosil gels~\cite{bellini-sil-PRE}.  The 
anisotropy factor is found to be $\alpha \approx 0.15$ in the two cases.  
For larger densities of aligned aerosil, the XY Bragg glass form is less 
successful in describing the measured line shape, as illustrated by 
Fig.~7 which shows the result of a fit using Eq.~(2) to the low-
temperature scattering intensity for $\rho_s$ = 0.10 g/cm$^3$.  We 
interpret this discrepancy as an indication that the larger aerosil 
densities impose positional disorder that is too strong for the XY Bragg 
glass to be stable.

\begin{figure}
\centering\includegraphics[scale=0.45]{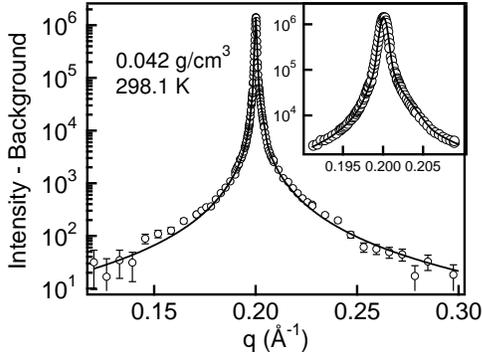}
\caption{Comparison between the background-subtracted scattering 
intensity as a function of wave vector for $\rho_s = 0.042$ g/cm$^3$ at 
298.1 K and the form predicted for an XY Bragg glass with a truncation 
in the correlations at large length scales.  The circles are the 
measured intensity, and the solid line is the result of a fit to 
Eq.~(2).  The inset shows a comparison between the scattering intensity 
and the fit result in the peak region on an expanded scale.  The XY 
Bragg glass form agrees closely with the measured intensity over the 
full range of wave vectors.} 
\label{XYBG0p042}
\end{figure}
\begin{figure}
\centering\includegraphics[scale=0.45]{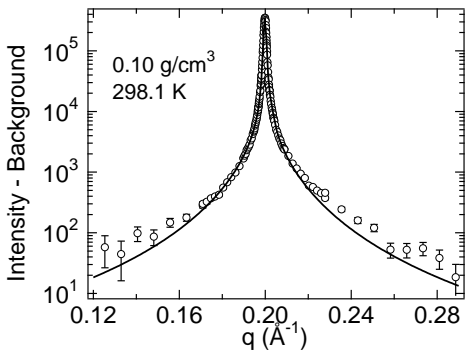}
\caption{Comparison between the background-subtracted scattering 
intensity as a function of wave vector for $\rho_s = 0.10$ g/cm$^3$ at 
298.1 K and the form predicted for an XY Bragg glass with a truncation 
in the correlations at large length scales.  The circles are the 
measured intensity, and the solid line is the result of a fit to 
Eq.~(2).  Deviations between the XY Bragg glass form and the measured 
intensity are apparent at large wave vectors.} 
\label{XYBG0p10}
\end{figure}

For the lower aerosil densities, the close agreement with the XY Bragg 
glass form extends over a range of temperatures below T$^*$.  However, 
at temperatures sufficiently close to T$^*$, contributions to the 
scattering from smectic critical fluctuations become appreciable.  These 
contributions are apparent in the two temperatures slightly below T$^*$ 
included in Fig.~4(a) where both smectic critical fluctuations at  large 
$(q_z-q_0)$ and additional contributions at smaller wave vector are 
apparent.   Efforts to account for the contributions from dynamic 
fluctuations in this temperature region by adding a term to Eq.~(2) are 
successful in the sense that good agreement with the measured scattering 
intensity can be achieved.  However, this analysis leads to large 
uncertainties in the fit parameters; therefore, we will not discuss this 
temperature region and will restrict our comparison to the XY Bragg 
glass form to lower temperatures, where the dynamic fluctuations can be 
safely neglected.  Figure 8 displays the values of the power-law 
exponent $\delta$ extracted from fits over this low-temperature region 
for $\rho_s$ = 0.027 and 0.042 g/cm$^3$.  The measured exponents for 
both densities are remarkably close to the universal value predicted for 
the XY Bragg glass, $\delta = 2.45$~\cite{toner_prl,saundersthesis}, 
shown by the dashed line in the figure.  We note that like the XY Bragg 
glass phase, the smectic phase of pure liquid crystals possesses 
quasi-long range order due to the Landau-Peierls instability and hence also 
displays a power-law singular scattering peak.  However, the power-law 
exponent along $q_{\|}$ for the smectic phase is always less than 
2~\cite{kaganer,als-nielsen}, a value that is incompatible with the 
measured line shapes for 8CB confined to aligned aerosil gels.  In 
contrast, the close quantitative agreement between the exponent 
predicted for the XY Bragg glass and the observed values lends strong 
support to the conclusion that this topologically ordered phase forms in 
8CB confined to aligned aerosil gels with low density.

\begin{figure}
\centering\includegraphics[scale=0.45]{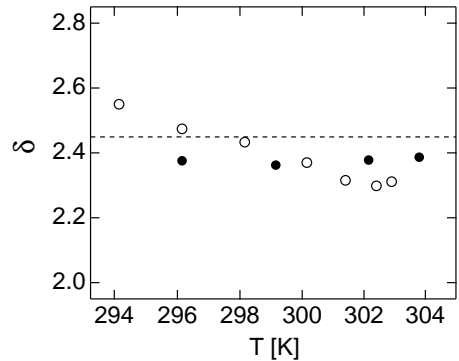}
\caption{Power-law exponent $\delta$ for the XY Bragg glass as a 
function of temperature for 8CB confined to aligned aerosil gels with 
densities $\rho_s = 0.027$ g/cm$^3$ (solid) and 0.042 g/cm$^3$ (open).  
The dashed line is the universal value for the exponent, $\delta = 
2.45$, predicted theoretically~\cite{toner_prl,saundersthesis}.}
\label{XYBGpower}
\end{figure}

Despite this excellent, quantitative agreement between the measured 
scattering profiles for low densities of aerosil and the XY Bragg glass 
form, we note that some caution is in order.  As described previously, 
one can analyze the scattering intensity using Eq.~(1) with  
$\xi_{2\|}=\xi_{1\|}$ and $\xi_{2\bot}=\xi_{1\bot}$.  (Although, the 
agreement is not nearly as accurate as that of the XY Bragg glass form 
nor does it cover as large a wave-vector range; see Fig.~5(c) in 
Ref.~\cite{Liang}).  Further, the values that are obtained for 
$\xi_{\|}$ and $\xi_{\bot}$ from fitting to Eq.~(1) are similar to those 
found for 8CB confined to isotropic aerosil gels following the same 
analysis~\cite{Liang}.  As we argue above, this analysis provides a 
flawed physical picture; however, this coincidence in the correlation 
lengths suggests that the smectic correlations that form in the aligned 
gels and the isotropic gels are similar.  Since the XY Bragg glass phase 
is not expected to be relevant to smectics confined to isotropic gels, 
one might interpret this similarity as evidence against the XY Bragg 
glass phase forming in the aligned gels.  Additional experimental work 
on other smectic liquid crystals confined to aligned gels that test for 
the presence of the XY Bragg glass phase would help in clarifying the 
stability of this topologically ordered phase within the anisotropic 
random environment created by the gel.
 
\section{High-Temperature Correlations:  Pseudo-critical Behavior}

As described above, the high-temperature scattering intensity of 8CB 
confined to aligned aerosil gels is characterized by dynamic critical 
fluctuations.  We interpret these fluctuations as the remnants of the 
N-SmA critical point that is destroyed by the quenched disorder (and not 
directly with the transition to the XY Bragg glass phase which 
presumably has a significantly subtler signature in the x-ray scattering 
than the observed robust peaks illustrated in Fig.~3).  Although the 
N-SmA critical point is obtained strictly only in the presence of zero 
quenched disorder, upon cooling in weak disorder the system in some 
sense comes in close proximity to the critical point and should 
therefore display critical smectic fluctuations.  Further, if the 
disorder is sufficiently weak, the susceptibility and correlation 
lengths charactering these fluctuations, while non-singular, should 
display pseudo-critical behavior over an extended range of reduced 
temperature.  This premise leads us to define the effective transition 
temperature T$^*$ as the temperature at which these quantities would 
diverge if their growth were not truncated:
\begin{eqnarray} 
\label{ch4_q0L}
q_0 \xi_{1\|} \sim |t|^{ -\nu_{||}}\\
\label{ch4_q0L2}
q_0 \xi_{1\bot} \sim |t|^{ -\nu_\bot}\\
\label{ch4_gamma}
\sigma_1 \sim |t|^{ -\gamma}
\end{eqnarray} 
where $t \equiv ($T$-$T$^*)/$T$^*$ is the reduced temperature, and 
$\nu_{\|}$, $\nu_{\bot}$, and $\gamma$ are effective critical exponents 
for $\xi_{1\|}$, $\xi_{1\bot}$, and $\sigma_1$, respectively.  

If this assumption of truncated power-law divergences is valid, then the 
three critical parameters should yield self-consistent values for T$^*$.  
To test this assumption and to identify the correct value of T$^*$, we 
fit the values of $\xi_{1\|}$, $\xi_{1\bot}$ and $\sigma_1$ to 
Eqs.~(\ref{ch4_q0L})-(\ref{ch4_gamma}).  In these fits, only data at 
high temperatures where the dynamic term in Eq.~(1) alone describes 
accurately the measured line shapes were included.  As mentioned above, 
at lower temperature (near but below T$^*$), additional contributions to 
the scattering intensity, presumably from the static correlations of the 
XY Bragg glass phase, appear.  Since efforts to separate these different 
contributions lead to considerable uncertainties in the fit parameters, 
we focus on the high temperature side of the critical region.  In the 
fitting, the critical exponents were held fixed and T$^*$ was treated as 
a free parameter.  These fits were repeated over a range of critical 
exponents, and the quality of the fits, as determined from the $\chi^2$ 
values, was compared.  For all aerosil densities, the minima in $\chi^2$ 
for the three power-law relations, Eqs.~(\ref{ch4_q0L})-(\ref{ch4_gamma}), 
occur at values of T$^*$ that are within 0.01 K of 
each other.  We equate T$^*$ with the average of the three values 
determined in this way for each aerosil density.  This value is listed 
in Table~\ref{ch4_table_result}.  Comparing T$^*$ for different aerosil 
densities, we note the value displays a non-monotonic dependence on 
$\rho_s$ with a minimum near $\rho_s = 0.05$ g/cm$^3$, consistent with 
the effective critical temperature determined with calorimetry for 8CB 
confined to isotropic aerosil gels~\cite{germano}.
\begin{table}
\caption{Summary of effective critical exponents for 8CB confined to 
aligned aerosil gels in comparison with pure 8CB and the 3D XY model.  
Also included are the gel pore chords $l_0$~\cite{germano} and effective 
transition temperatures T$^*$. }
\vspace{10pt}
\centering\begin{tabular}{|c||c|c|c|c|c|c|}
\hline
System&$l_0$ (\AA)&T$^*$ (K)&$\gamma$&$\nu_{\|}$&$\nu_{\bot}$&$\nu_{\|} 
/ \nu_{\bot}$\\[3pt]
\hline
\hline
Pure 8CB&&306.97&1.26&0.67&0.51&1.314\\[3pt]
\hline
\multirow{2}{*}{0.027 g/cm$^3$}&\multirow{2}{*}{$\sim 
3000$}&\multirow{2}{*}{305.96}&1.55&0.73&0.70&\multirow{2}{*}{1.05}\\[3pt]
&&&$\pm$0.05&$\pm$0.02&$\pm$0.02&\\[3pt]
\hline
\multirow{2}{*}{0.042 g/cm$^3$}&\multirow{2}{*}{$\sim 
1500$}&\multirow{2}{*}{305.01}&1.54&0.74&0.65&\multirow{2}{*}{1.13}\\[3pt]
&&&$\pm$0.05&$\pm$0.03&$\pm$0.03&\\[3pt]
\hline
\multirow{2}{*}{0.071 g/cm$^3$}&\multirow{2}{*}{$\sim 
900$}&\multirow{2}{*}{305.62}&1.56&0.75&0.66&\multirow{2}{*}{1.15}\\[3pt]
&&&$\pm$0.05&$\pm$0.02&$\pm$0.02&\\[3pt]
\hline
\multirow{2}{*}{0.10 g/cm$^3$}&\multirow{2}{*}{$\sim 
600$}&\multirow{2}{*}{305.86}&1.53&0.73&0.65&\multirow{2}{*}{1.13}\\[3pt]
&&&$\pm$0.05&$\pm$0.02&$\pm$0.03&\\[3pt]
\hline
3D XY&&&1.32&0.67&0.67&1\\[3pt]
\hline
\end{tabular}
\label{ch4_table_result}
\end{table}
\begin{figure}
\centering\includegraphics[scale=0.22]{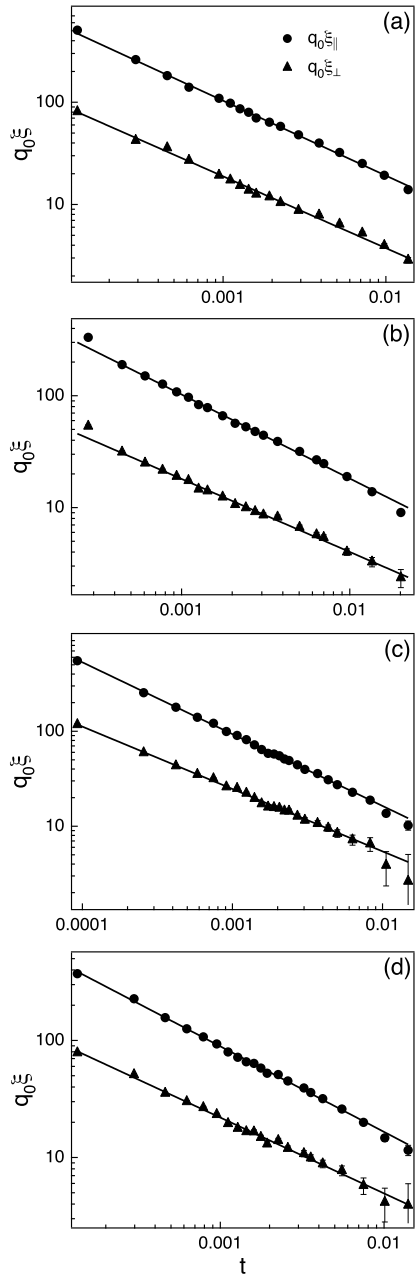}
\caption{Dynamic correlation lengths $\xi_{1\|}$ (circles) and 
$\xi_{1\bot}$ (triangles) as a function of reduced temperature for 8CB 
confined to aligned aerosil gels with densities (a) $\rho_s = 0.027$ 
g/cm$^3$, (b) $\rho_s = 0.042$ g/cm$^3$, (c) $\rho_s = 0.071$ g/cm$^3$, 
and (d) $\rho_s = 0.10$ g/cm$^3$.  The solid line are the results of 
power-law fits.}
\label{ch4_corr_t}
\end{figure}
\begin{figure}
\centering\includegraphics[scale=0.22]{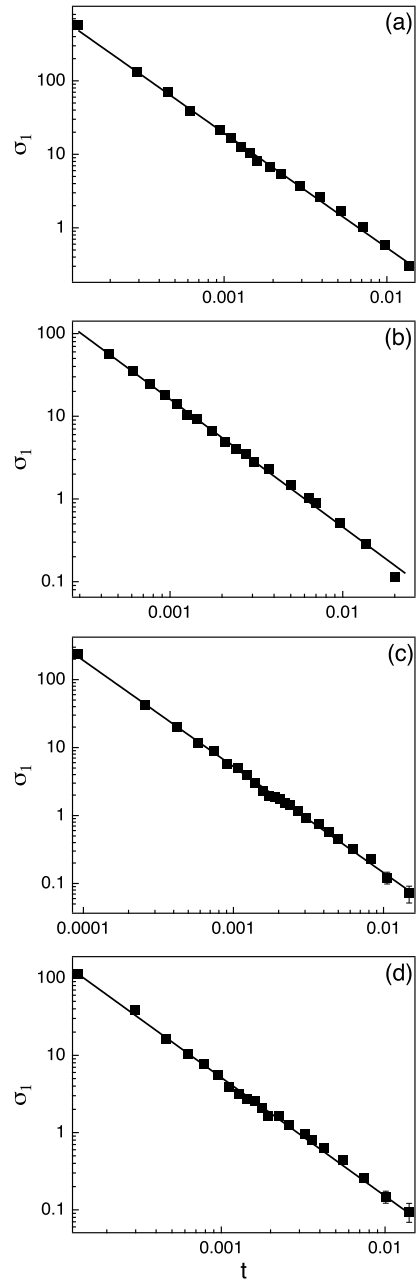}
\caption{Smectic susceptibilty as a function of reduced temperature for 
8CB confined to aligned aerosil gels of density for (a) $\rho_s = 0.027$ 
g/cm$^3$, (b) $\rho_s = 0.042$ g/cm$^3$, (c) $\rho_s = 0.071$ g/cm$^3$, 
and (d) $\rho_s = 0.10$ g/cm$^3$.  The solid line are the results of 
power-law fits.}
\label{ch4_s1_t}
\end{figure}
Figures \ref{ch4_corr_t} and \ref{ch4_s1_t} display log - log plots of 
$\xi_{1\|}$, $\xi_{1\bot}$ and $\sigma_1$ as a function of reduced 
temperature for T $>$ T$^*$ for various gel densities using the optimal 
values of T$^*$.  The solid lines in the figures are the results of fits 
to Eqs.~(\ref{ch4_q0L})-(\ref{ch4_gamma}), which demonstrate that the 
temperature dependence is well described by power laws, further 
reinforcing the idea of pseudo-critical behavior.
The effective critical exponents extracted from the fits are listed in 
Table~\ref{ch4_table_result} along with the values for pure 
8CB~\cite{ocko,J_theon,davidov}.  The values of $\nu_{\|}$, 
$\nu_{\bot}$, and $\gamma$ for 8CB confined to aligned aerosil gels are 
distinctly different from those of pure 8CB, indicating that the 
quenched disorder introduced by the gel not only affects the 
low-temperature smectic order but also modifies the pre-transitional 
critical fluctuations.  At sufficiently high temperature the correlation 
lengths are small compared to the mean pore chord $l_0$ of the 
gels~\cite{germano}, which is also listed in 
Table~\ref{ch4_table_result}. 
Hence, the disorder caused by the gels should have little direct effect 
on the smectic fluctuations well above T$^*$.  In the simplest picture 
of the N-SmA transition, the formation of the smectic density wave 
breaks 3D XY symmetry, and thus the critical behavior of pure liquid 
crystals could be expected to match that of the 3D XY model.  However, 
the observed critical behavior of pure 8CB and other smectic liquid 
crystals typically deviates from that of the 3D XY model.  These 
deviations are illustrated by the difference in critical exponent values 
for pure 8CB and the 3D XY model listed in Table~\ref{ch4_table_result}.  
The precise nature of these deviations, whose source likely involves 
couplings between the smectic order parameter and the nematic order, has 
been a long-standing problem.  
As described above, when a smectic liquid crystal is confined to an 
aligned aerosil gel, it experiences not only positional disorder that 
couples to the smectic density wave but also anisotropic orientational 
fields that couple to the nematic order, 
thereby modifying the nematic behavior.  We therefore conclude that 
differences in the critical fluctuations between pure 8CB and 8CB 
confined to aerosil gels originate from the coupling of nematic order to 
the gel and its indirect influence on the smectic critical behavior. 

\begin{figure}
\centering\includegraphics[scale=0.45]{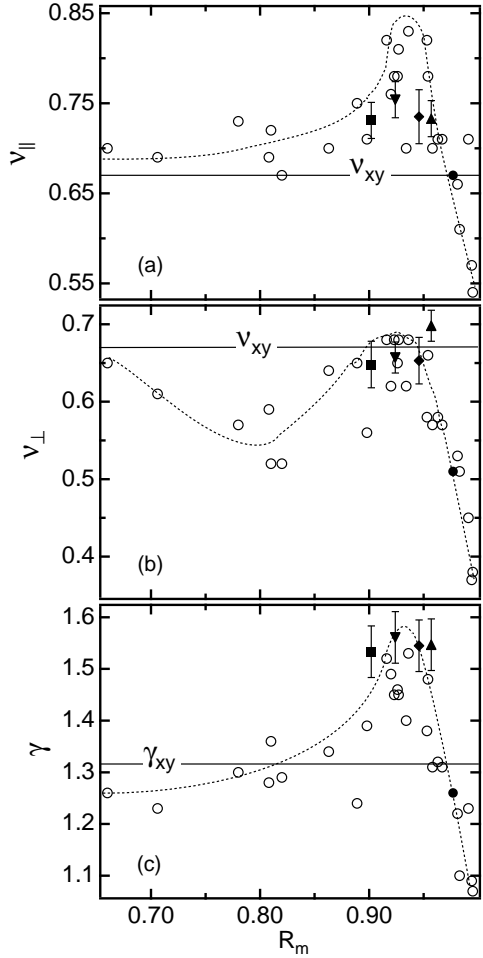}
\caption{Critical exponents (a) $\nu_{\|}$, (b) $\nu_{\bot}$, and (c) 
$\gamma$ for different smectic liquid crystals as a function of McMillan 
ratio, $R_M \equiv \frac{T_{NA}}{T_{NI}}$.  The values for various pure 
liquid crystals, shown by circles, are taken from~\protect\cite{nounesis}.  The 
values for pure 8CB are shown by the solid circles.  Also included are 
the values of the effective critical exponents for 8CB confined to 
aligned aerosil gels with densities $\rho_s = 0.027$ g/cm$^3$, 
(triangle), $\rho_s = 0.042$ g/cm$^3$ (diamond), $\rho_s = 0.071$ 
g/cm$^3$ (upside-down triangle), and $\rho_s = 0.10$ g/cm$^3$ (square), 
which are assigned effective McMillan ratios using 
Eq.~(\ref{ch4_Rm_eff}).  The solid lines mark the exponent values for 
the 3D XY model, and the dashed lines are guides to the eye.}
\label{ch4_Rm}
\end{figure}

As an illustration, compare the ratio $\nu_{\|} / \nu_{\bot}$ for 
varying aerosil densities with the ratio for pure 8CB listed in 
Table~\ref{ch4_table_result}.  For the 3D XY model, the ratio is one due 
to the isotropic scaling of correlations.  However, for pure 8CB the 
scalings are highly anisotropic ($\nu_{\|} / \nu_{\bot}$ = 1.314), 
consistent with behavior seen in various other smectic liquid 
crystals~\cite{nounesis}.  Patton and Andereck have advanced a theory 
that this anisotropic scaling results from coupling between nematic 
director fluctuations and the smectic order parameter~\cite{andereck, 
patton}.  
For 8CB confined to aligned aerosil gels, the ratio 
$\nu_{\|}/\nu_{\bot}$ shows much weaker scaling anisotropy than in pure 
8CB.  One interpretation for this decreased scaling anisotropy is that 
pinning of the nematic director by the anisotropic orientational fields 
suppresses long-wavelength nematic fluctuations, thus weakening their 
influence over the smectic critical behavior.  

While the decrease in $\nu_{\|}/\nu_{\bot}$ for 8CB confined to aligned 
aerosil gels compared with $\nu_{\|}/\nu_{\bot}$ for pure 8CB indicates 
qualitatively that the quenched disorder weakens nematic-smectic 
coupling, this observation would be made stronger with a quantitative 
measure of this weakening.  In pure liquid crystals, the strength of the 
nematic-smectic coupling can be crudely parameterized by the McMillan 
ratio 
\begin{equation}
R_M \equiv \frac{T_{NA}}{T_{NI}}
\end {equation}
specifying the temperature range of the nematic phase.  Large $R_M$ 
indicates a short nematic range, which typically implies the nematic 
order is far from saturated at the N-SmA transition.  Hence, for large 
$R_M$ the nematic susceptibility can be expected to be large, and 
concomitantly the order parameter coupling can be expected to be strong.  
In a comprehensive survey of smectic critical behavior in pure liquid 
crystals, Garland and Nounesis found that the critical exponents 
$\alpha$, $\gamma$, $\nu_{\|}$, and $\nu_{\bot}$ obtained from 
calorimetry and x-ray scattering studies displayed complex but 
systematic trends as a function of $R_M$~\cite{nounesis}.  Figures 
\ref{ch4_Rm}(a)-\ref{ch4_Rm}(c) display the values of the critical 
exponents $\nu_{\|}$, $\nu_{\bot}$, and $\gamma$ of various smectic 
liquid crystals as a function of $R_M$, as originally assembled by 
Garland and Nounesis.  
Liquid crystals with small $R_M$ generally have critical exponents that 
approach the 3D XY values; however, the values of the exponents do not 
change monotonically with decreasing $R_M$.  The critical exponents for 
pure 8CB ($R_M$ = 0.977) are shown by the solid circles in the figures.  
Also in Figs.~\ref{ch4_Rm}(a)-\ref{ch4_Rm}(c) are the effective critical 
exponents for 8CB confined to aligned aerosil gels, which have been 
placed on the figures by assigning the smectics with quenched disorder 
an effective McMillan ratio,
\begin{equation}
R_M^{eff} = 0.977 - 0.72 \rho_s
\label{ch4_Rm_eff}
\end{equation}
As the figures illustrate, this mapping between disorder strength, as 
parameterized by aerosil density, and nematic-smectic coupling, as 
parameterized by McMillan ratio, places the effective critical exponents 
for 8CB confined to aligned aerosil densities roughly in line with 
trends of pure liquid crystals of varying $R_M$.  We note a very similar 
linear mapping of aerosil density to effective McMillan ratio was 
introduced previously for 8CB and 4O.8 confined to isotropic aerosil 
gels~\cite{paperI,Larochelle}.  
In these cases, the mapping was shown to collapse successfully effective 
heat capacity and order parameter exponents, $\alpha$ and $\beta$, for 
various aerosil densities onto the trends of pure liquid crystals.  The 
linear coefficient that optimized the mapping for the smectics confined 
to isotropic aerosil gels was smaller than that in 
Eq.~(\ref{ch4_Rm_eff}), 0.47 as opposed to 0.72, suggesting that aligned 
aerosil gels are more effective than isotropic gels in suppressing 
nematic-smectic order parameter coupling.  Nevertheless, these linear 
mappings provide a quantitative measure of how the presence of the 
quenched disorder influences the smectic critical behavior.  
The results in Figs.~\ref{ch4_Rm}(a)-\ref{ch4_Rm}(c) thus extend the 
notion of an effective McMillan ratio to the effective critical 
exponents $\nu_{\|}$, $\nu_{\bot}$, and $\gamma$, making comprehensive 
the analogy between disorder strength and $R_M^{eff}$ for describing the 
behavior of the pre-transitional critical fluctuations of smectic 
confined to aerosil gels.  Further theoretical efforts to understand the 
effects of quenched disorder could hence be a fruitful avenue for 
unraveling the outstanding mysteries surrounding the critical behavior 
of pure smectics.

\vspace{1 cm}
\noindent
{\bf Acknowledgements:}
We gratefully acknowledge C. Garland, G. Iannacchione, B. Ocko, and L. 
Radzihovsky for helpful discussions.  Funding was provided by the NSF 
under Grant No.~DMR-0134377.  Use of the National Synchrotron Light 
Source, Brookhaven National Laboratory, was supported by the U. S. 
Department of Energy, Office of Science, Office of Basic Energy 
Sciences, under Contract No.~DE-AC02-98CH10886.


 
 
\end{document}